\documentclass{aa} 
\usepackage{graphics}
\usepackage{psfig}

\def\msun{M$_{\sun} $}
\def\rxs{1RXS J012851.9--233931}

\begin{document}

\newcommand{\rb}[1]{\raisebox{1.5ex}[-1.5ex]{#1}}

\thesaurus {06(02.01.2; 08.14.2; 08.09.2 1RXS J012851.9--233931; 08.13.1)}

\title{Zeeman lines and a single cyclotron line in the low-accretion rate polar
1RXS J012851.9--233931 (RBS0206)
\thanks{Based on observations at the 
European Southern Observatory La Silla (Chile)}
}
\author {A.D. Schwope\inst{1}
        \and 
           R. Schwarz\inst{1} 
        \and
           J. Greiner \inst{1}
}
\offprints{Axel D.~Schwope}
\mail{ASchwope@aip.de}
\institute {
        Astrophysikalisches Institut Potsdam, An der Sternwarte 16,
        14482 Potsdam, Germany
}
\date{Received 31 March 1999, accepted}
\titlerunning{Zeeman lines and a single cyclotron line in RBS0206}
\maketitle

\begin{abstract}
We present a low-resolution discovery spectrum and 
CCD-photometry of the bright X-ray source 
1RXS J012851.9--233931 found in the ROSAT All-Sky Survey.
These first observations suggest that the source is an AM Herculis star
(polar) accreting at a low rate. The optical spectrum is dominated by  
Zeeman absorption features from the white dwarf, indicating a mean photospheric
magnetic field of $36\pm 1$\,MG. Only weak Balmer line emission was
observed. In the near infra-red, a single 
intense cyclotron hump was observed. The inferred magnetic field strength
in the accretion plama is $45\pm1$\,MG, the temperature in the plasma
is below 2\,keV.
Likely orbital periods are $\sim$90\,min or $\sim$146\,min, 
the latter inside the cataclysmic variable period gap. 
The system is an ideal target for further detailed investigations 
of the field structure on a magnetic white dwarf by phase-resolved 
spectropolarimetry. 
\end{abstract}


\section{Introduction}
Several dozen new AM Herculis binaries (polars) were identified in recent 
years as optical counterparts of bright soft X-ray/EUV emitters, most 
of them in the systematic identification program of bright, soft, high-galactic
latitude X-ray sources found in the RASS (ROSAT All-Sky Survey) 
by Thomas et al.~(1998). Other main sources of new systems are the 
ROSAT/WFC survey (Pye et al.~1995) and serendipitous ASCA/ROSAT/EUVE 
discoveries.
Thus, a total of about 60 polars are known meanwhile, more than three times
the number of sources from the pre-ROSAT era. New interesting individual 
systems, e.g.~bright eclipsing systems, have emerged and systematic
studies became possible concerning e.g.~the period or magnetic field 
distribution.


We are running an identification program of all bright, high-galactic latitude
sources found in the RASS, primarily in order to establish complete X-ray 
selected samples of extragalactic X-ray emitters. Selection criteria are a
RASS count rate above 0.2\,s$^{-1}$ and galactic latitude $|b| > 30\degr$.
This program, termed the 
ROSAT Bright Survey RBS (Fischer et al.~1998), has impact also 
on galactic work. Among other fields, we mention the quest for isolated 
neutron stars (Schwope et al.~1999, Neuh\"auser \& Tr\"umper 1999).
By expanding the selection criteria applied by Thomas et al.~including fainter 
and harder sources we also find new cataclysmic variables. The first one 
emerging from this program, RBS1735 (=EUVE\,J2115--58.6), 
turned out to belong to the 
rare class of only 4 polars with a slightly asynchronously rotating white 
dwarf (Schwope et al.~1997). This system had, contrary to the large majority 
of the newly discovered 
polars a rather hard X-ray spectrum. 
Here we report on the discovery 
of a new polar with a soft X-ray spectrum. The new system shows two features
of magnetic activity, photospheric Zeeman absorption features and a cyclotron 
emission spectrum. Hence, it is a prime candidate for a detailed investigation 
establishing a map of the magnetic field on the surface of the white 
dwarf. The present database is too small to achieve this and further 
observations (photometry, spectroscopy, spectro-polarimetry) at optical, 
infra-red and X-ray wavelength are highly demanded.

\section{Observations}
\subsection{X-ray observations}
\rxs\ was detected during the RASS as a 
variable soft X-ray source at a mean countrate of 0.34 \,s$^{-1}$. The source
was scanned 23 times during the RASS and had a total exposure of 452 sec.
It displayed flux variations by 100\% with minimum countrate of zero and 
peak countrate of 0.93\,s$^{-1}$. 
A variability analysis revealed no obvious periodicity.
It has a soft spectrum with X-ray hardness ratio
HR1 = $-0.84 \pm 0.04$, where HR1 is defined in the usual way, 
HR1 $= (H - S)/(H + S)$ with $H$ and $S$ being the counts in the soft 
(0.1--0.4\,keV) and hard (0.5--2.0\,keV) bands, respectively.
The X-ray source appeared as object 206 in the target list of the RBS 
(Schwope et al.~1999, in preparation), we refer to it as RBS0206 in the 
following.

It was not detected in the all-sky surveys performed in the softer 
spectral regimes with the EUVE satellite (Bowyer et al.~1996, Lampton 
et al.~1997)
and with the WFC onboard ROSAT (Pye et al.~1995).

\subsection{Low-resolution spectroscopy}
\begin{figure}
\psfig{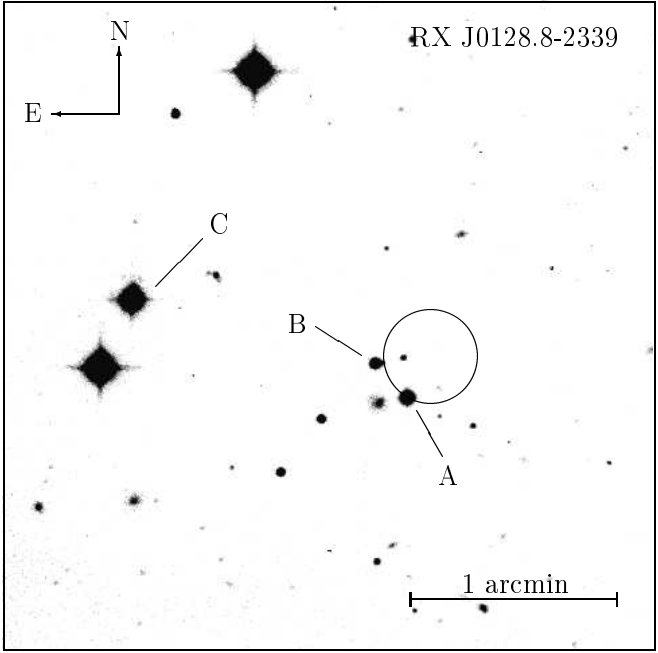}
\caption{Finding chart for RBS0206, the image is the average of all Gunn $i$
CCD frames of Jan.~24. The image size is
3 by 3 arcminutes. The X-ray source, an 
AM Herculis star, is object `A'. The optical coordinates are 
$\alpha_{\rm 2000} = 01^h28^m52\fs3, \delta_{\rm 2000} = -23\degr39'44''$}
\label{f:chart} 
\end{figure}

The X-ray positional uncertainty of RBS0206  as given in the 1RXS-catalogue 
(Voges et al.~1996) is 9\,arcsec. The digitized sky survey DSS shows 
several possible counterparts at or just within the 90\% confidence X-ray error
circle. The object nearest to the RASS X-ray position at a distance of 
$7.6$\arcsec\ is the faintest (Gunn $i = 19\fm9$), it 
appears just above the plate limit and was not observed optically by us.
The others labeled `A' and `B' on the finding chart reproduced in 
Fig.~\ref{f:chart} are at distances of $13.2$\arcsec\ and $15.9$\arcsec, 
respectively.

\begin{figure*}[t]
\psfig{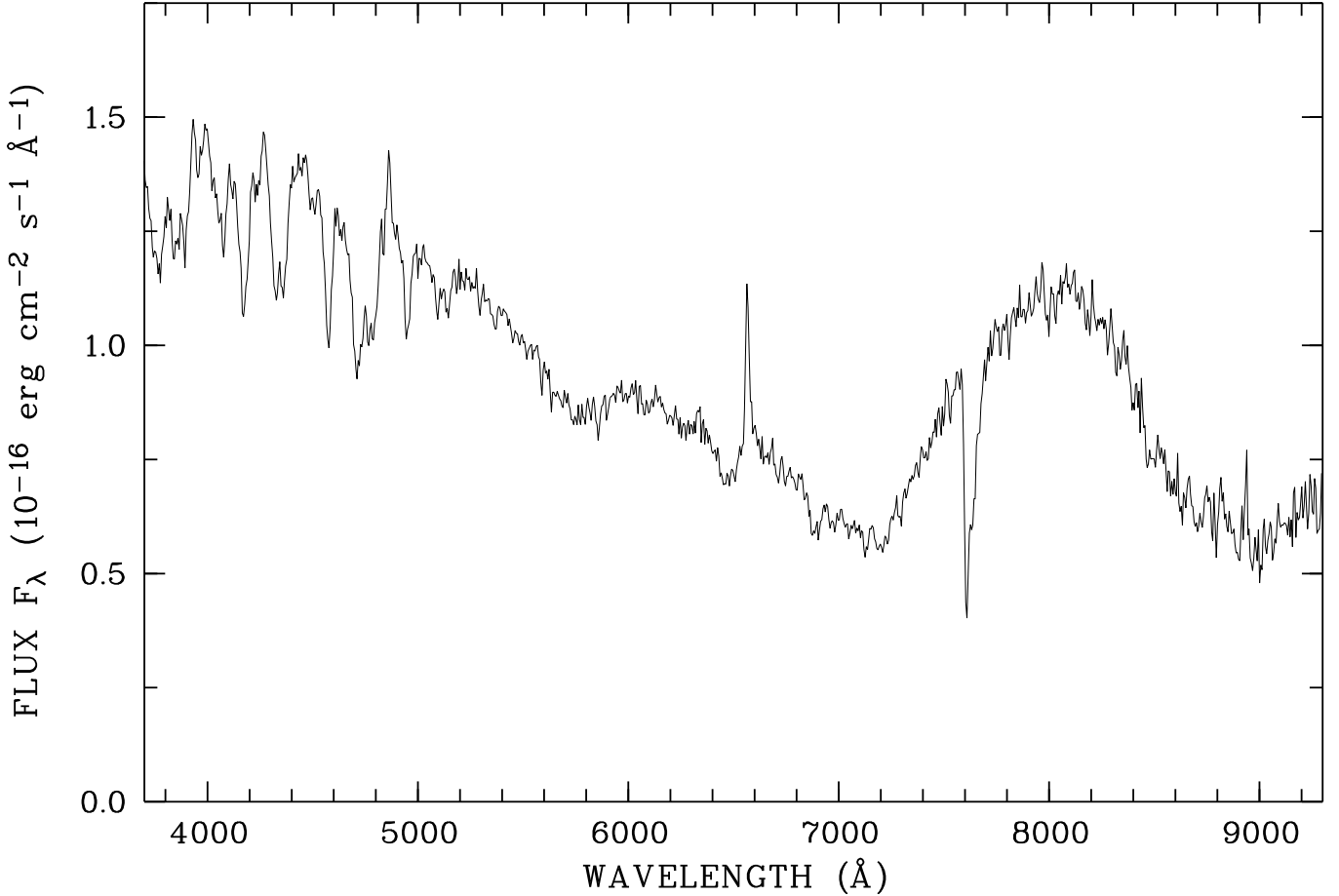}
\caption{Low-resolution discovery spectrum of \rxs\ (=RBS0206) taken with the 
ESO 3.6m telescope and EFOSC2 on November 14, 1998}
\label{f:spec} 
\end{figure*}

Low-resolution spectra of objects `A' and `B' 
were taken with the ESO Faint Object Spectrograph and Camera EFOSC 
mounted to the ESO 3.6m telescope in the night November 14, 1998.
Both objects were put on the spectrograph slit, the integration time was
1200 sec, the exposure started at 4.2872 UT. A grism with 
236 grooves/mm was used as dispersive device (grism \#13), 
providing a wavelength 
coverage of more than 5000\,{\AA} at a resolution of about 12\,{\AA} FWHM.
Spectrophotometric standard stars were observed twice in the night. Due to 
transparency changes of the atmosphere these observations gave inconsistent 
photometric results. 
The observations of the 
standard stars were nevertheless 
useful in order to establish the overall spectral response.
An absolute calibration of the stars in the field was possible with the 
CCD-photometry of Jan.~1999 (see next section). 
We estimate the finally achieved 
photometric accuracy of our spectra to be better than $\sim$15\,\%,
where this estimate is based on the photometric accuracy of our 
Jan.~1999 imaging observations.
Object `B' turned out to be a faint K-star ($V = 19\fm1$) and is 
not the counterpart of the 
RASS X-ray source. The spectrum of object `A' is reproduced in 
Fig.~\ref{f:spec}. With $V = 18\fm9$ this star was slightly brighter than 
`B'. It has a blue continuum with deep absorption lines  
and an intense broad hump centred on $\sim$8000\,{\AA}. H$\alpha$ was 
observed to be  weak in 
emission. The object was identified as a likely AM Herculis star in a low
state of accretion. 

\subsection{Optical photometry}
\begin{figure}[t]
\psfig{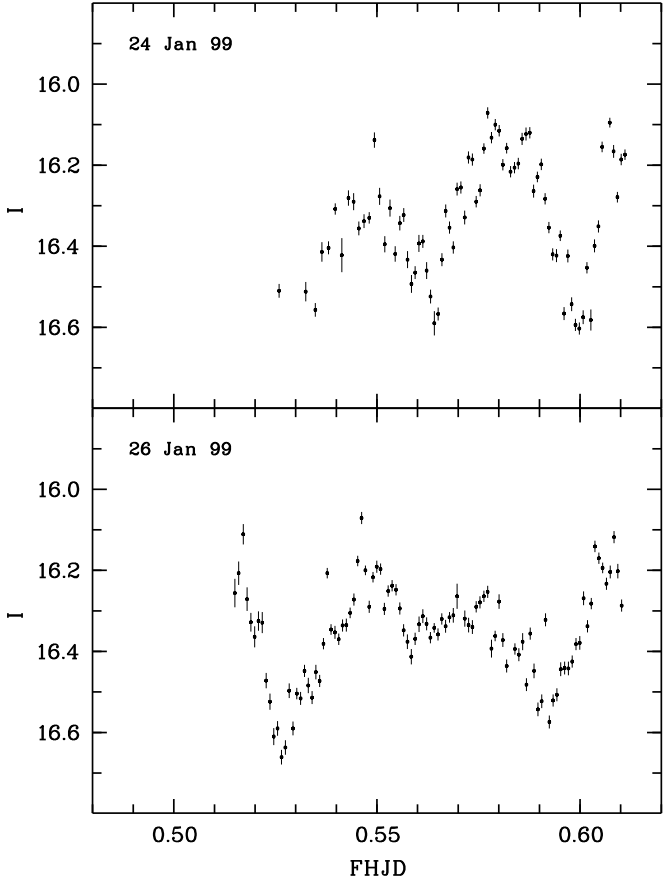}
\caption{Time-resolved optical photometry of RBS0206  through Gunn $i$ filter. 
The abscissa gives fractional heliocentric Julian days with corresponding 
integer parts 2451202 and 2451204, respectively}
\label{f:lc_ori} 
\end{figure}

Optical photometry of RBS0206  was obtained 
on January 24 and 26, 1999, 
with the Danish 1.5m telescope at ESO La Silla. 
The source was exposed for a total of
2 and 2.3 hours, respectively, with integration times of individual 
exposures of 60\,sec. Main aim of these observations was the determination 
of the orbital period of the binary. The feature with expected highest 
photometric variability in the 
spectrum of RBS0206  (Fig.~\ref{f:spec}) is the hump at 8000\,{\AA}. 
The photometric observations were therefore performed with a Gunn $i$ 
filter with central wavelength at 7978\,{\AA} \ and width 
of 1430\,{\AA}.
Differential magnitudes of objects `A' and `B' 
with respect to star `C' were derived using the profile fitting algorithm 
of DOPHOT (Mateo \& Schechter 1989). 
The accuracy of a single observation is $0.05$\,mag.
Subsequent observations of Landolt (1992) standards allowed the absolute
calibration of our photometric and spectroscopic reference stars,
$i = 13\fm92$ for star `C' and $i = 17\fm38$ for star `B'.
The resulting light curves of the two occasions are shown in 
Fig.~\ref{f:lc_ori}. RBS0206 displayed brightness variations 
by 0.5\,mag on a timescale of less than one hour with short-term fluctuations
of 30\% superimposed. The mean brightness level of $i \sim 16\fm3$ indicates
a brightening of the source with respect to the spectroscopic observations
by about $1.3$\,mag.

\section{Analysis and discussion}
\subsection{Spectroscopic identification of RBS0206}
The spectrum of RBS0206  shows the signs of a polar in a low state of 
accretion.
The emission lines of Hydrogen and Helium, which are so prominent in the 
high accretion states of these systems, have almost vanished. Only H$\alpha$
and H$\beta$ are weakly detectable 
in emission. The absence of all radiation components dominating 
the optical spectra of polars in their high states (atomic emission lines, 
recombination continuum, quasi-continuous cyclotron emission) unveils 
the presence of the magnetic white dwarf in RBS0206. 
The continuum in the blue spectral range is dominated by the 
white dwarf, the pronounced absorption lines are Zeeman shifted Balmer 
lines of photospheric origin (H$\alpha$, H$\beta$ and H$\gamma$). 
Although in a low state, the spectrum has a broad intensity hump in the 
near-infrared at 8000\,{\AA} \ and width of about 1000\,{\AA} \ (FWHM) which 
can be nothing else than a cyclotron feature. Hence, accretion did not cease
entirely. The spectrum does not show any obvious feature from the 
mass-donating secondary star. We will discuss the various spectral 
features subsequently, but try to determine the binary period beforehand.

\subsection{Optical variability}
\begin{figure}[t]
\psfig{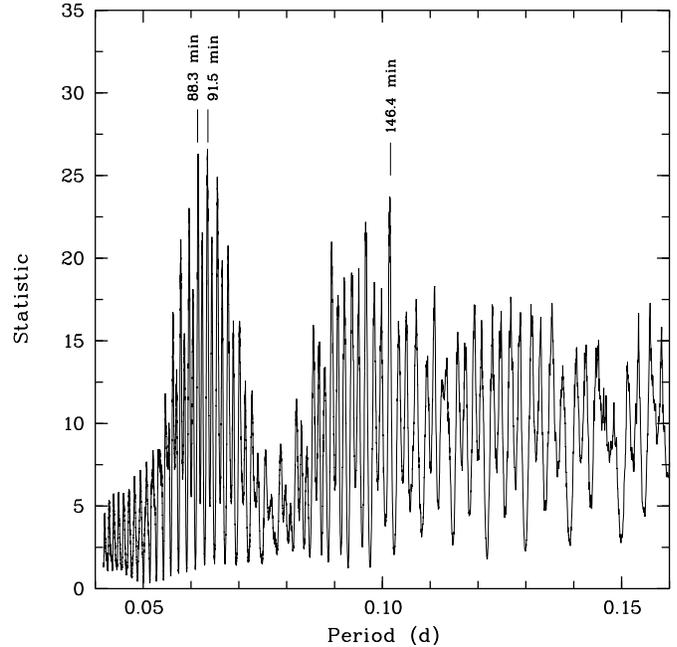}
\caption{Periodogram (AOV-statistic) of the time-resolved data shown in 
Fig.~\ref{f:lc_ori}}
\label{f:periodo} 
\end{figure}

\begin{figure}[t]
\psfig{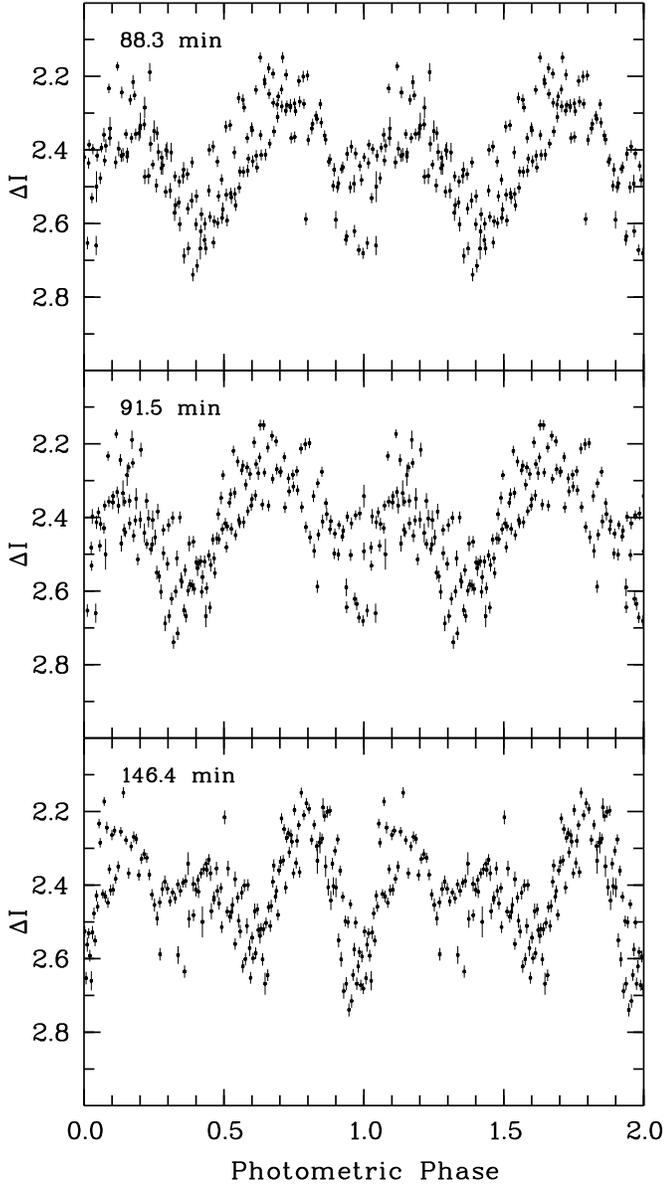}
\caption{Folded light curves using trial periods indicated in 
Fig.~\ref{f:periodo}}
\label{f:lc_fold} 
\end{figure}

The lightcurves shown in Fig.~\ref{f:lc_ori} display pronounced minima
which are in principle useful for a period determination.
With the small database at hand, however, our results are not unique.
It is in particular not clear, whether the two minima in the lightcurve
of Jan.~26, 1999, represent the same feature in two successive 
cycles or not. A periodogram
computed with the {\it analysis-of-variance (AOV)} algorithm 
(Fig.~\ref{f:periodo}; Schwarzenberg-Czerny 1989)
reveals possible 
periods around 90\,min (0.0625 days) and around 146\,min (0.1 days). 
The former solution would imply that the two observed minima on Jan.~26
represent the same feature, the latter that they 
are different. Folded light curves using the three trial periods 
indicated in Fig.~\ref{f:periodo} are shown in Fig.~\ref{f:lc_fold}. 
Each folded light curve
has intervals with a very large scatter of data points at given phase
probably caused by cycle-to-cycle variations of the brightness.
The folded light curves for periods around 
90\,min resemble those of MR~Ser or QQ~Vul, which both have their 
main accreting pole continously in view. The optical light curves 
of these two systems are modulated by strong cyclotron beaming. Primary 
minima occur in these systems when we are looking almost pole-on, secondary
minima in the centre of the bright hump, when the main pole vanishes 
partly behind the limb of the white dwarf. 
The folded light curve using  the trial period of 146.4\,min
shows a rather short bright phase (centred on phase zero) which is 
disrupted by an eclipse-like minimum. Such a feature may be caused by 
absorption in the intermittent accretion stream crossing  the line
of sight to the accretion spot. 
We regard a period around 90\,min as likely but cannot exclude longer periods, 
significantly shorter periods are certainly excluded. Clearly, observations
with a longer time base are needed to clarify the situation.
With this 
period, the likely accretion geometry is such, that the accretion spot
is continously in view. 
This view is supported by the RASS X-ray light curve which has non-vanishing 
X-ray flux for all but one scans.

\subsection{Spectral analysis}
The absence of any obvious M-star feature in our spectrum 
can be used to derive a distance estimate to the system. 
We assume that the maximum contribution of the M-star at 9000\,{\AA} is 80\% 
and use as template spectrum for the secondary in RBS0206  the M6 dwarf
Gl406 with $M_K = 9\fm19, V-K = 7\fm37$. This type of M-star would be 
appropriate for a cataclysmic binary with $P_{\rm orb} \sim 90$\,min.
The scaled $V$- and $K$-band 
brightnesses of Gl406 are $V_{\rm sc} = 23\fm5$ and $K_{\rm sc} = 16\fm1$.
We assume a Roche-lobe filling secondary star at a 
period of 90\,min which has a spherical equivalent Roche radius of 
$\log(R_2/R_{\sun}) = -0.86$.
Using Bailey's (1981) method combined with the improved calibration of 
the surface brightness of late-type stars 
by Beuermann \& Weichhold (1999, in prep.) which predicts a surface brightness 
$S_K = 4.9$ for a star like Gl406, the  distance to RBS0206 
is $> 240$\,pc.

Using the observed slope and flux level of the white dwarf in the blue
spectral regime the questions of white dwarf radius, temperature and
distance to the system can be addressed, too. For that exercise we 
use the model spectra for non-magnetic white dwarf atmospheres by 
G\"{a}nsicke et al.~(1995) kindly provided by B.~G\"{a}nsicke.
We assume a normal 0.6\,M$_{\sun}$ white dwarf with $R_{wd} = 8 \times 
10^8$\,cm.
The observed spectrum is reasonably well reflected with a 10000\,K white dwarf 
at a distance of only 130\,pc, although the model predicts a steeper 
spectral slope than observed. A white dwarf at a distance of 240\,pc
as estimated above must have a considerably higher temperature of about 
20000\,K in order to match the observed flux level at 6000\,{\AA}. 
At this high temperature the continuum slope is much steeper than observed
and the fit in general is clearly worse than that for 10000\,K.
In order to resolve the discrepancy between the 
different distance estimates one definitely 
needs phase-resolved data. These would allow in the first place 
to determine the orbital 
period. Should our period estimate for some reason be wrong and the 
orbital period shorter than 90 min, one can hide an even fainter 
secondary star with later spectral type 
in the spectrum which would be less distant than the 
derived 240\,pc for a period of 90\,min. Phase-resolved spectroscopic data
in the blue spectral regime would allow to disentangle between radiation 
from the undisturbed photosphere and a warm accretion spot. 

\begin{figure*}[t]
\psfig{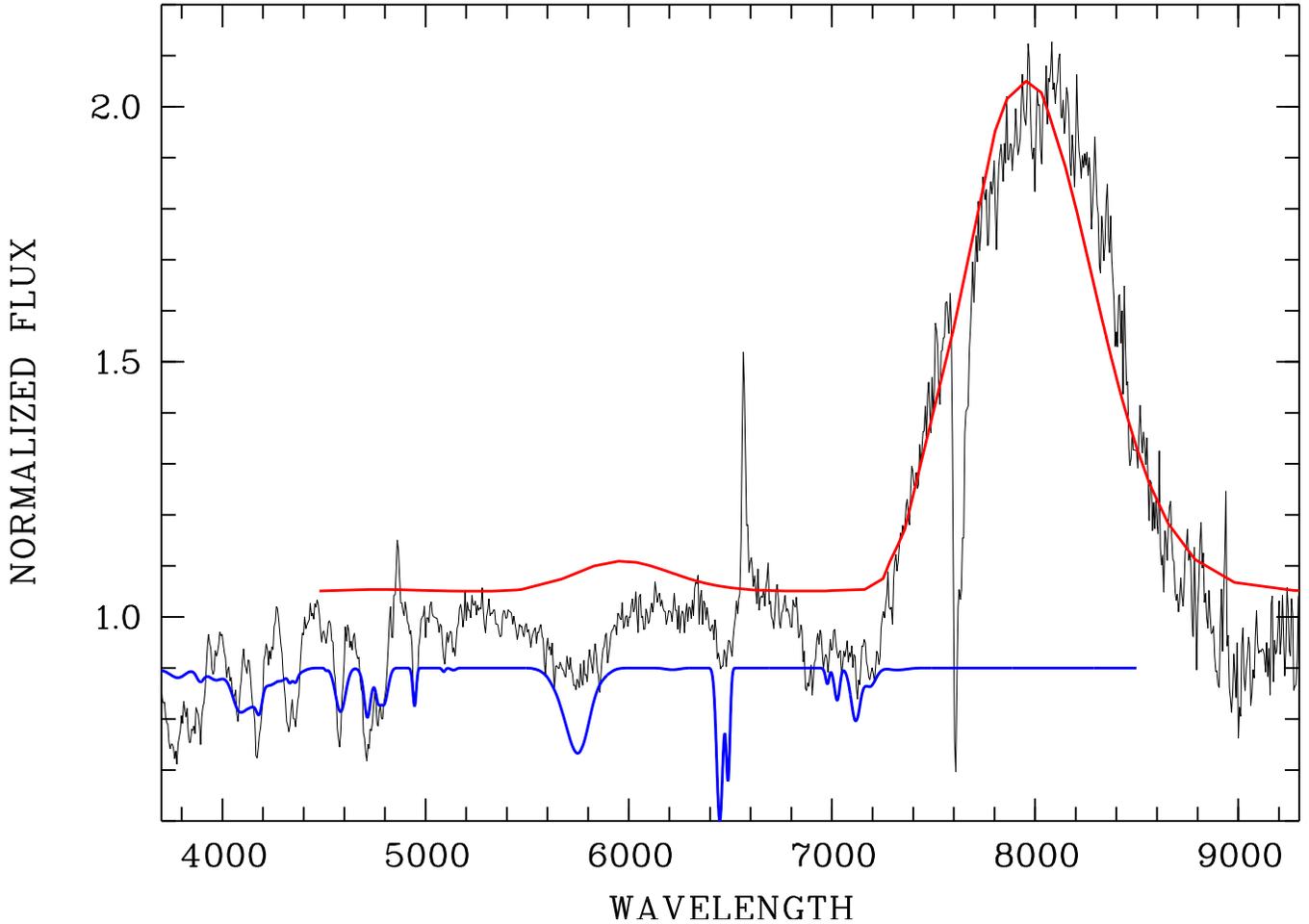}
\caption{Normalized spectrum of the white dwarf in RBS0206  together 
with simple Zeeman and cyclotron models used for the determination 
of the mean magnetic field and the field in the accretion spot}
\label{f:zeemod} 
\end{figure*}

The purity of the Zeeman spectrum shortward of H$\alpha$ allows a 
direct measurement of the mean magnetic field strength. We fitted a
second order polynomial to interactively defined continuum points and 
divided the observed spectrum by this curve. 
The result is shown in Fig.~\ref{f:zeemod}
together with a simple Zeeman model plotted below the observed spectrum. 
The model is based on the detailed computations of the wavelengths and
oscillator strengths of the individual non-degenerate Zeeman transitions 
of H-Balmer lines by Forster et al.~(1984) and R\"{o}sner et al.~(1984).
For the present purpose we use the transitions 
of H$\alpha$, H$\beta$ and H$\gamma$,
which lie in the spectral range covered by our spectrum.
Our model just sums the oscillator strengths of all Balmer transitions 
mentioned weighted according to an assumed magnetic field distribution.
For the model shown in Fig.~\ref{f:zeemod} we assumed a Gaussian 
field distribution centred on $B = 36$\,MG with a spread $\sigma_B = 2$\,MG. 
The model, therefore, does not predict real spectral intensities but
it predicts the wavelengths where spectral features are expected to occur.
At the field strength realized in RBS0206 nearly 
all subcomponents of the Balmer lines appear as individual non-degenerate 
transitions due to the dominance of the quadratic over the linear
Zeeman effect.
Due to magnetic field smearing and to the limited spectral resolution 
these cannot be resolved in our spectrum, the Zeeman lines mainly appear as
broad troughs. All features longward of 5300\,\AA\ belong to H$\alpha$, 
the features shortward of this wavelength belong to H$\beta$ and H$\gamma$ and
become partly intermixed.
There are, however, some isolated features e.g.~at 4580\,\AA\
and 4920\,\AA\ reacting sensitively on the adopted value of the 
centroid field strength.
We estimate the uncertainty of the centroid field strength to be 
about 1\,MG. We regard this field strength as mean photospheric field strength.
If we want to infer the value of the field strength 
at the pole, we need to know (a) the inclination of the magnetic axis with 
respect to the line of sight at the time of the observation, (b) the true
underlying distribution of the magnetic field, and (c) information 
about the temperature distribution over the white dwarf surface. 
All these quantities and distributions are unknown. However, if we assume
a dipolar field structure, a homogeneous temperature distribution
and a line of sight
not directly towards the magnetic pole, then 
a typical value for the conversion 
between the mean magnetic field and the polar field strength 
is 1.6 -- 1.7. Using these values
a likely value of the magnetic field at the pole is 55 -- 60\,MG. A 
lower limit for the field strength at the magnetic pole 
is given by the high-field wings of the observed Zeeman lines, 
which indicate about 40\,MG.

The occurence of a single cyclotron hump at 8000\,{\AA} is puzzling. 
All AM Herculis stars with individually resolved cyclotron harmonics
show more than one harmonic in their optical or infrared spectra. 
One apparent exception is the recently discovered polar HS\,1023+3900
(Reimers et al.~1999) observed at a very low accretion rate
which shows at certain phases only one prominent
cyclotron line at 6000\,{\AA}. 
Only the detailed analysis revealed a second cyclotron line
as a faint addition to the bright spectrum of the 
secondary star in the near infrared at 9000\,{\AA}. 
In RBS0206  the cyclotron fundamental lies in the unobserved
infrared, $\lambda_{\rm cyc} = 17850 - 26780$\,{\AA} \ for
$B_{\rm pole} \simeq B_{\rm cyc} = 40 - 60$\,MG.
Our spectrum covers the corresponding 
harmonic numbers 2 -- 5.3 or 3 -- 8, respectively,
hence we could expect to observe several higher cyclotron harmonics.
The only way to hide these higher harmonics is 
to assume that the particular observed 
one is already almost optically thin, that the 
plasma temperature is very low (which gives a steep dependence of the 
absorption coefficient on the harmonic number) and that 
the plasma is rather dilute. As one example for such a model 
we show also in Fig.~\ref{f:zeemod}
a cyclotron model which fits these requirements. It is computed for a 
homogeneous, isothermal plasma with $kT = 2$\,keV, magnetic field strength
$B = 45$\,MG, viewing angle $\Theta = 50\degr$ (angle between magnetic
field and observer) and plasma density parameter $\log \Lambda = 2$ (see 
Barrett \& Chanmugam 1985, Thompson \& Cawthorne 1987). With these parameters
the observed cyclotron hump is the $3^{\rm rd}$ harmonic, the $2^{\rm nd}$
is expected to lie in the infrared J-band (centred on 1.2\,$\mu$m) and the
cyclotron
fundamental lies at 2.4\,$\mu$m, at the edge of the K-band. The next higher 
harmonic, the fourth, is seen in the model as a small hump at 6000\,{\AA}.
Such a feature
can easily be overlooked in somewhat noisy data and can be hidden in 
a continuum which is
strongly affected by Zeeman absorption. Our modeled  hump at 8000\,{\AA}\
has the same width as the observed one. 
Since any inhomogeneity in the emitting plasma like a variation of the
value of the magnetic field strength tend to smear a cyclotron line, the 
true plasma temperature is probably below 2\,keV.

This is the lowest plasma temperature among all polars derived so far 
from optical cyclotron spectroscopy.
It is in particular much lower than the shock temperature of a
free-falling accretion stream on the white dwarf, 
by a factor of 10 for an assumed 0.6\,M$_{\sun}$ white dwarf.
This suggests an accretion scenario termed `bombardement solution' (Kuijpers
\& Pringle 1982) where the accretion spot is heated by particle collisions.
This scenario has been worked out in detail in a series
of papers by Woelk \& Beuermann (1992, 1993, 1996) and 
Beuermann \& Woelk (1996). Interpreting the measured temperature as 
maximum electron temperature, Figs.~6, 8, and 9 of their 1996 paper
suggest, that the specific mass accretion rate $\dot{m}$ is below 
0.1\,g\,cm$^{-2}$\,s$^{-1}$.
Their modelling predicts furthermore 
that 100\% of the accretion luminosity appears
as cyclotron radiation. This is in apparent contradiction to the fact,
that RBS0206 was discovered as X-ray source. However, AM Herculis binaries 
are known to have highly variable mass accretion rates on different 
time scales and we cannot exclude something like a high accretion state 
during the RASS and a low accretion state during our spectroscopic 
observations.

The very low value of the plasma parameter $\Lambda$
supports the applicability of the bombardement solution and predicts an
even lower specific mass accretion rate. The plasma parameter 
is given in terms of the magnetic field strength $B_{45} = B/45\,MG$, 
the geometric path 
length through the plasma $s_6 = s/10^6$\,cm 
and the electron density $n_{16} = n_e/10^{16}$\,cm$^{-3}$ as
$\Lambda = 1.3 \times 10^6 s_6 n_{16} B^{-1}_{45}$. With the measured value of 
$\Lambda = 100$ this becomes $n_{16} \sim 10^{-4} s_6^{-1}$. 
The cooling length $h$ of the plasma will not be smaller than $10^6$\,cm 
(Beuermann \& Woelk 1996, Fig.~2), and by setting the cooling length $h$ equal 
to the geometrical path length $s$ the electron density in the cyclotron 
region must be extraordinarily low, $n_e \simeq 10^{12}$\,cm$^{-3}$. For 
standard composition the post-shock density is $n_e = 3.8 \times 10^{17}
(\dot{m}/10^2 {\rm g\,cm}^{-2} \\{\rm s}^{-1})(M_{\rm wd}/M_{\sun})^{-1/2}%
(R_{\rm wd}/10^9 {\rm cm})$\,cm$^{-3}$, suggesting $\dot{m} \simeq 10^{-3}$ for
the case of RBS0206. 

The predicted integrated cyclotron flux (integrated over all harmonics)
is about $2 \times 10^{-12}$\,erg\,cm$^{-2}$\,s$^{-1}$. 
For a distance of $130$\,pc the cyclotron luminosity is 
$L_{\rm cyc} = \pi d^2 F_{\rm cyc} \\ = 1 \times 10^{30} 
(d/130)^2$\,erg\,s$^{-1}$.
Assuming that 100\% of the accretion luminosity is released as cyclotron 
radiation, as predicted by the Woelk \& Beuermann models, the total 
mass accretion is $\dot{M} = LR_{\rm wd}/GM_{\rm wd} = 1 \times 10^{13} 
\mathrm{g/s} = 1.5\times 10^{-13}$\,\msun yr$^{-1}$. 
This value is $2 -3 $ orders of magnitude below the canonical value
for a short-period polar, i.e.~a system below the period gap,
in a high accretion state.

The RASS X-ray data can be fitted with a soft blackbody and a (marginally 
detected) hard X-ray brems\-strah\-lung model. The integrated unabsorbed 
flux of the $kT_{\rm bb} = 20$\,eV blackbody gives 
$5 \times 10^{-11}$\,erg\,cm$^{-2}$\,s$^{-1}$, which is a factor of 25
more than the derived cyclotron flux. These estimates
suggest that the accretion rate has changed by a large amount between 
the RASS X-ray observations and the optical spectroscopy in November 1998.
  
\section{Conclusions}
We have presented first spectroscopic and photometric observations of 
the newly discovered AM Herculis star \rxs\ = RBS0206. Although not finally 
conclusive, our photometry suggests that it is a short-period system
with $P_{\rm orb} \simeq 90$\,min or that it is a system in the period
gap with $P_{\rm orb}$ around 140\,min. The probable accretion 
geometry is such, that the accreting pole is continuously in view.
The discovery spectrum of RBS0206  was taken when the system was in 
an extreme low state of accretion. This derives from the absence 
of bright emission lines and photospheric absorption lines from the 
white dwarf. Even in the single discovery spectrum 
Zeeman and cyclotron lines could be detected. The inferred magnetic field
strengths are $36\pm1$\,MG for the mean 
photospheric field $45\pm1$\,MG for the accretion spot. The plasma temperature 
in the cyclotron emission region is extremely low, $kT < 2$\,keV,
compared to the more usual 5 -- 20\,keV encountered in most AM Her systems.
The derived integrated mass accretion rate is 2 -- 3 orders of magnitude 
below the canonical value for short-period polars indicating a deep low 
accretion state of that system.
RBS0206 seems to be an extraordinarily good target for a detailed study 
of the magnetic field structure over the white dwarf surface 
due to the presence of pronounced Zeeman lines 
and a cyclotron line in addition. 
The Zeeman lines are sensitive to the average surface field
whereas the cyclotron line goves an extra constraint on the field 
in one particular spot. The system is also an excellent target for studies 
of cyclotron-line formation in the lowest harmonics by spectroscopy
in the J and K bands, where pronounced 
optical depth effects are expected to occur (Woelk \& Beuermann 1996).

\begin{acknowledgements}
We thank Klaus Beuermann (G\"{o}ttingen) for his surface brightness 
calibration of late-type stars and Boris G\"{a}nsicke (G\"{o}ttingen) 
for providing a grid of white dwarf atmosphere models.

This work was supported by the German 
Bun\-des\-mi\-ni\-sterium f\"ur Bildung, Wissenschaft, Forschung 
und Technologie (BMBF/DLR) under grants 50 OR 9403 5, 50 OR 9708 6 and 
50 QQ 9602 3.

The ROSAT project is supported by BMBF/DLR and the Max-Planck-Society.
\end{acknowledgements}

\end{document}